\documentclass{article}
\usepackage[]{graphicx}
\usepackage[]{epsfig}

\newcommand{\beq}{\begin{equation}}
\newcommand{\eeq}{\end{equation}}
\newcommand{\beqa}{\begin{eqnarray}}
\newcommand{\eeqa}{\end{eqnarray}}
\newcommand{\non}{\nonumber}
\renewcommand{\a}{\alpha}
\newcommand{\D}{\Delta}

\newcommand{\arctanh}{{\rm arctanh}}

\begin{document}

\title{On the Complexity of the Bethe Lattice Spin Glass}
\author{Tommaso Rizzo
\\
\small Center for Statistical Mechanics and Complexity, INFM Roma ``La Sapienza'' 
\\
\small Piazzale Aldo Moro 2, 00185 Roma, 
Italy}

\maketitle

\begin{abstract}
A theory for the complexity of the Bethe lattice spin-glass is developed applying to the cavity-method scheme of M\'ezard and Parisi
the results recently obtained in the context of the Sherrington-Kirkpatrick model.
The crucial ingredient is the introduction of a new  cavity field $z$ 
related to the marginality of the relevant states.
The theory admits a variational formulation.
In the high-connectivity limit it yields the Bray and Moore expression for the TAP complexity of the SK  model.
An annealed version of the theory is also studied in order  to obtain non-trivial results at low computational cost.
An analysis of the theory is performed
numerically through population-dynamics algorithms and 
analytically through power series expansion.
The results can be applied to other finite connectivity problems.
\end{abstract}

\section{Introduction}

The work of M\'ezard and Parisi on the Bethe lattice spin-glass \cite{MP1,MP2} has shown that the cavity method is an essential tool in the context of disordered system.
Conceived originally in order to recover the results of the ``replica method without replicas'' \cite{MPVc}, its application  to finite connectivity problems and in particular the possibility of applying population-dynamics algorithms to obtain non-perturbative results
has given a considerable boost to the field of disordered systems and optimization problems in the last three years, see {\it e.g.} \cite{MZ,MZP}.  
In this work we will apply the results recently obtained  in the context of the SK model \cite{ABM,PR,R} to the Bethe lattice spin glass, developing a general theory for the logarithm of the number of states of this model, {\it i.e.} the Complexity.
The theory can be applied also to other finite connectivity and optimization problems.    
We will basically follow the path opened in \cite{MP1} whose detailed knowledge is therefore essential to read this work. 
For this reason we will not present a self-contained discussion of the Bethe lattice spin-glass and
the notation as well will be borrowed from \cite{MP1}.

\subsection{Basic Ideas of the Theory}
\label{sec:bas}

The basic idea of the theory is that the relevant states at a given 
  temperature, 
{\it i.e.} those yielding the total complexity must be marginal states . Indeed if 
  we expect the complexity to decrease upon an infinitesimal increase of  the 
  temperature these states must suddenly disappear \cite{R}.
This is indeed what happens in the SK model according to the BM theory due 
  to the presence of a vanishing  isolated eigenvalue \cite{ABM,PR,R} in the TAP 
  spectrum.
This makes the application of the Cavity method to these states impossible 
  because one of the basic hypotheses of the method is that the states 
  can be continued analytically upon adding a new spin to the system.
In \cite{R} it was shown how to avoid this problem in the context of the SK 
  model and the whole BM theory was recovered within the cavity method. The 
  idea  is to consider states different from the relevant 
  ones. This can be done for instance considering the number of states with a self-overlap $q$ different from  $q^*$, the self-overlap of the relevant states. 
These non-relevant states exist in an exponential number although smaller than the relevant ones.
These states are no more marginal and the Cavity method can be applied to them in order to compute their complexity; at the end the result for the total complexity is obtained by taking the limit $q\rightarrow q^*$. In this limit the marginality of the states reflects itself in the divergence of some parameters of the theory. In this work we use these ideas to develop a theory for the complexity of the spin-glass on the Bethe lattice.

In \cite{R} the selfoverlap $q$ of the TAP solutions was used in order
 to select a set of solutions different from the relevant ones, here we stress that {\it any   parameter $X$ but the free energy can be used}. Indeed the total complexity cannot depend on the parameter used and in the limit $X\rightarrow X^*$ the computations with different parameters are equivalent. 
In the present context we will use the magnetization and not the selfoverlap in order to take full advantage of the tree structure of the lattice. Indeed the relevant objects to be computed are the shifts of the quantities of interest in various branch-merging processes and in the next section we will see that the shift of the magnetization, being simply the  derivative of the free energy shift, can be expressed in terms of a few branch parameters, {\it i.e.} the cavity fields $h_i$ and their derivatives $h_i'$. The same computation with the self-overlap would be much more complicated.
Therefore we will compute the function 
$\tilde{\Sigma}(u,\lambda_M)$ which in the thermodynamic limit is the Legendre transform of the complexity $\Sigma(f,M)$ 
\beq
\Sigma(f,M)=\ln \sum_{\alpha} \delta(M-M_{\alpha}) \delta (f-f_{\alpha})\ \ \ ;\ \ \ 
\tilde{\Sigma}(u,\lambda_M)=\ln \sum_{\alpha}e^{-u \beta f_{\alpha}-\lambda_M M_{\alpha}}
\label{deffe}
\eeq
In particular we are interested in the $\lambda_M\rightarrow 0$ limit of $\tilde{\Sigma}(0,\lambda_M)$ that yields the total complexity.

\section{The Cavity Method on the Bethe Lattice}

\subsection{Extensive Quantities}
\label{subsec:theext}

In this subsection we report two important relationships that allow us to express an extensive quantity in terms of averages of proper shifts. Such relationships can be proved using the topological argument represented pictorially in fig. (2) of  \cite{MP2}.  
If the quantity $A$ is extensive, that is 
\beq
\lim_{N\rightarrow \infty}A=N a \, ,
\eeq
then the following relationships holds
\beq
a= {k+1\over 2}\langle \Delta A^{(2)} \rangle -k \langle \Delta A^{(1)} \rangle 
\label{exte1}
\eeq
\beq
a={k+1\over 2}\langle \Delta A^{(iter)} \rangle -{k-1\over 2} \langle \Delta A^{(1)}\rangle
\label{exte2}
\eeq
where $\Delta A^{(2)} $ is the shift of $A$ upon adding two spins to a system of $2k$ branches, $\Delta A^{(1)}$ is the shift of $A$ upon adding a spin to a system of $k+1$ branches and $\Delta A^{(iter)}$ is the shift upon adding a spin to a system of $k$ branches.
The square brackets mean disorder average.
Following \cite{MP2} we note that
the first relationship can be obtained removing $2k$ spin from the system $S_1$ that is a Bethe lattice of $N$ spin. The new system is called $S_0$ and has $N-2k$ spins. Then the  system $S_2$ is obtained  merging the $2k(k+1)$ free branches of $S_0$ in groups of $2k$ to a couple of spins. Now $S_2$ is a system of  $N+2$ spins and 
\beq
a={1\over 2}\langle A(S_2)-A(S_1) \rangle ={1\over 2}\langle (A(S_2)-A(S_0))-(A(S_1)-A(S_0)) 
\eeq
where the square brackets mean disorder average.  Now equation (\ref{exte1}) can be  easily derived noting that 
\beq 
\langle A(S_2)-A(S_0) \rangle =(k+1)\langle \Delta A^{(2)}\rangle\ \ \ ;
\langle A(S_1)-A(S_0) \rangle =2 k\langle \Delta A^{(1)}\rangle\ \ \ ;
\eeq
The second relationship, eq. (\ref{exte2}) can be obtained in a similar way. We start removing $k$ spin from the system $S_1$ obtaining the system $S_0$ with $N-k$ spins. Then the $k(k+1)$ free branches are merged in groups of $k$ yielding the system $S_{int}$ with $N+1$ spins and $k+1$ free branches. Merging the $k+1$ branches of $S_{int}$ we obtain the system $S_{fin}$ that is a system with the topology of the Bethe lattice ({\it i.e.} with no free branches) and $N+1$ spins. Then the shift between the system $S_{fin}$ and $S_{1}$ can be expressed in terms of $\langle \Delta A^{(1)}\rangle$ and $\langle \Delta A^{(iter)}\rangle$ yielding eq. (\ref{exte2}).

\subsection{Fields, Shifts and Distribution Functions}

The function $\tilde{\Sigma}(u,\lambda_M)$ is an extensive quantity, therefore according to the previous section it can be expressed as the average over the disorder of proper shifts. The shift in $\tilde{\Sigma}$ in the process of merging if $k+1$ branches is given by:
\beq
\Delta \tilde{\Sigma}(u,\lambda_m)^{(1)}=\ln {\sum_{\a} e^{u \beta F_{\alpha}+\lambda_m M_{\alpha}+u \beta \Delta F_{\alpha}^{(1)}+\lambda_M \Delta M_{\alpha}^{(1)}} \over\sum_{\a} e^{u \beta F_{\alpha}+\lambda_m M_{\alpha}}}
\label{defdf}
\eeq
The free energy and magnetization shifts can be computed starting from the following equation \cite{MP1}:
\beq
\sum_{\sigma= \pm 1}\exp(\beta \sigma_0 J \sigma+\beta H_{ext}\sigma_0+\beta h \sigma)=c(J,h)\exp (\beta u(J,h)\sigma_0+\beta H_{ext}\sigma_0)
\eeq
where 
\beq
u(J,h)={1\over \beta}\arctanh [\tanh (\beta J)\tanh (\beta h)]\ \ ;\ \  c(J,h)=2{\cosh(\beta J)\cosh (\beta h)\over \cosh (\beta u(J,h))}
\eeq
The free energy shift in the process of merging $k+1$ lines is given by \cite{MP1}:
\beq
-\beta \D F^{(1)}=\ln \left[2 \cosh \left(\sum_{i=1}^{k+1}\beta u (J_i,h_i)+\beta H_{ext} \right) \right]+\sum_{i=1}^{k+1}\left[  {\cosh \beta J_i  \over \cosh  \beta u(J_i,h_i)} \right]
\label{df1}
\eeq
The magnetization shift can be computed deriving the previous expression with respect to the external field $H_{ext}$:
\beqa
\D M^{(1)} & = &- {d\over d H_{ext}}\D F^{(1)}=\tanh \left(\sum_{i=1}^{k+1}\beta u (J_i,h_i)+\beta H_{ext} \right) +
\non
\\
& + & \sum_{i=1}^{k+1}\left(  \tanh \left(\sum_{j=1}^{k+1}\beta u (J_j,h_j)+\beta H_{ext} \right) -\tanh \beta u(J_i,h_i)       \right){d u \over d h_i} h_i'
\label{dM1}
\eeqa
Notice that the magnetization shift depends on the fields $h_i$ but also on their derivatives 
\beq
h_i'={dh_i\over dH_{ext}}
\eeq 
this is the most important modification of the standard approach.
We also write the expressions of the new field $h_0$ and $h_0'$ acting on the central spin in the merging process; the equation for $h_0$ is the same of \cite{MP1} and  by derivation we obtain the one of $h_0'$
\beqa
h_0&=&\sum_{i=1}^{k+1}u(J_i,h_i)+H_{ext}
\label{h0}
\\
h_0'&=&\sum_{i=1}^{k+1}{du(J_i,h_i)\over dh_i}h_i'+1
\label{h0p}
\eeqa

The hypotheses of the Cavity Method \cite{MPVc,MP1} can  be easily extended to the present discussion. We assume that the free energies $F_{\a}$ and magnetization $M_{\a}$ of the states are independent random variables with respect to the disorder. It is also assumed that the free energy and magnetization of a state are independent of the field $h_{\alpha}$ and of its derivative $h_{\alpha}'$. 
The distribution is 
such that on each sample the average number of states with free energy $F$ and  magnetization $M$ is exponential. Therefore on a given sample we can write:
\beq
d {\mathcal N}(F,M,h,h')\propto \exp [u \beta F+\lambda_M M]  P_i(h,h')
\eeq  
Where $P_i(h,h')$ represents the fraction of states \cite{Mon} with fields $h$ and $h'$  and the suffix $i$ is used to remember that this function  depends on the branch considered.
These hypotheses are self-consistent under the process of merging of $k+1$ branches; indeed we have:
\beqa
d{\mathcal N}(F^{(k+1)},M^{(k+1)},h_0,h_0')& \propto & \int \exp[u \beta F+\lambda_M M]dFdM\prod^{k+1}_{i=1}P_i(h_i,h_i')dh_i dh_i' \times 
\non
\\
& \times & \delta \left(h_0-\sum_{i=1}^{k+1}u(J_i,h_i)-H_{ext}\right)\delta\left(h_0'-\sum_{i=1}^{k+1}{du(J_i,h_i)\over dh_i}h_i'-1\right) \times
\non
\\
& \times & \delta(F+\Delta F^{(1)}-F^{(k+1)})\delta (M+\Delta M^{(1)}-M^{(k+1)}) \propto
\non
\\
& \propto & \exp[\beta u F^{(k+1)}+\lambda_M M]P_0^{(k+1)}(h_0,h_0')
\label{grancav}
\eeqa 
From the last equation the expression for $P_0^{(k+1)}(h_0,h_0')$ can be obtained in terms of the distributions $P_i(h_i,h_i')$.
The suffix $(k+1)$ is used to distinguish the distribution of the fields $h_0$ and $h_0'$ coming from the process of merging $k+1$ branches from those of the fields $h_i$ and $h_i'$ that instead are obtained merging $k$ branches.
It is important to notice that while the final free energy and magnetization are correlated to the fields $h_i$ and $h_i'$ they are not correlated with the field $h_0$ and $h_0'$. Furthermore, since the free energy and magnetization before the merging process are not correlated to the fields $h_i$ and $h_i'$, they are not even correlated to the free energy and magnetization shifts; therefore in the summation in eq. (\ref{defdf}) the part depending on the fields factorizes and we obtain:
\beq
\Delta \tilde{\Sigma}(u,\lambda_m)^{(1)}=\ln \int \prod_{i=1}^{k+1}P_i(h_i,h_i')dh_idh_i'\ \exp[u \beta \Delta F^{(1)}+\lambda_M \Delta M^{(1)}]
\eeq 
In order to obtain the disorder average of the previous shift we need to know the distribution over the disorder of the functions $P(h_i,h_i')$. Since these functions refer to the system before adding the central spin they are not correlated and we can consider the distribution of $P_i(h_i,h_i')$ on a generical branch. This distribution of distributions satisfies a recursive equation. Let us start deriving the expression for $P_0(h_0,h_0')$ in the process of merging $k$ branches. By definition it is the distribution of the field $h_0$ and $h_0'$ when the states are weighed with a weight proportional to $\exp[\beta u F_{\alpha}+\lambda_M M_\alpha]$:
\beq
P_0(h_0,h_0')={\sum_{\a}e^{\beta u F_{\alpha}+\lambda_M M_{\alpha}}\delta(h_0-h_{\alpha})  \over\sum_{\a}e^{\beta u F_{\alpha}+\lambda_M M_{\alpha}}  }
\eeq
A recursive expression of $P_0(h_0,h_0')$ can be obtained either by expressing the previous equation in terms of the fields $h_i$ and $h_i'$ through the analogue of eqs. (\ref{h0}) and (\ref{h0p}) 
and recalling that the shifts are not correlated to the free energy and  magnetization before the merging or equivalently from  the analogue of  relation (\ref{grancav}) applied to the merging of $k$ branches. The result is:
\beqa
P_0(h_0,h_0')& =& 
\frac{1}{\int \prod_{i=1}^{k}P_i(h_i,h_i')dh_idh_i'e^{u \beta \Delta F^{(iter)}+\lambda_M \Delta M^{(iter)}}} \int \prod_{i=1}^{k}P_i(h_i,h_i')dh_idh_i' \times
\non
\\
& \times &
e^{u \beta \Delta F^{(iter)}+\lambda_M \Delta M^{(iter)}} 
\delta \left(h_0-\sum_{i=1}^{k}u(J_i,h_i)-H_{ext}\right)\delta\left(h_0'-\sum_{i=1}^{k}{du(J_i,h_i)\over dh_i}h_i'-1\right)
\label{Ph0h0p}
\eeqa
where $\Delta M^{(iter)}$ is obtained much as $\Delta M^{(1)}$, eq. (\ref{dM1}),  merging $k$ branches instead of  $k+1$. 
Then the distribution of distributions ${\mathcal P}[P(h,h')]$ satisfies the following equation:
\beq
{\mathcal P}[P(h,h')]=\int\prod_{i=1}^k {\mathcal P}[P(h_i,h_i')]dP_i(h_i,h_i') \langle \delta( (P(h,h')-P(h_0,h_0'))\rangle_J
\eeq
Where $P(h_0,h_0')$ depends on $P_i(h_i,h_i')$ through eq. (\ref{Ph0h0p}) and the square brackets mean average over the $k$ bonds $J_i$.

\section{The $\lambda_M\rightarrow 0$ Limit} 
In order to study the total complexity or more generally the curve $\Sigma(f)$ we must study the $\lambda_M\rightarrow 0$ limit of the theory. We will show that this is different from simply putting $\lambda_M=0$ everywhere in the theory if the fields $h_i'$ diverge in this limit. In order to account for this divergence we  introduce the rescaled fields $z_i$:
\beq
z_i=\lambda_M h_i'
\eeq
According to eq. (\ref{h0p}) $z_0$ satisfies the following recursion equation
\beq
z_0=\sum_{i=1}^{k}{du(J_i,h_i)\over dh_i}z_i+\lambda_M
\eeq
The theory can be simply reformulated in terms of the fields $z_i$ at $\lambda_M$ finite because the magnetization shift is always multiplied by $\lambda_M$ in the expression of the complexity shift or in the recursion equation for $P_0(h_0,h_0')$; therefore the relevant object is:
\beqa
\lambda_M \D M^{(1)} & = &=\lambda_M \tanh \left(\sum_{i=1}^{k+1}\beta u (J_i,h_i)+\beta H_{ext} \right) +
\non
\\
& + & \sum_{i=1}^{k+1}\left(  \tanh \left(\sum_{j=1}^{k+1}\beta u (J_j,h_j)+\beta H_{ext} \right) -\tanh \beta u(J_i,h_i)       \right){d u \over d h_i} z_i
\label{lambdaDM}
\eeqa
In the limit $\lambda_M\rightarrow 0$ the previous expression  remains finite if the fields $z_i$ remain finite. Notice that instead the first term will go to zero anyway.
The introduction of the fields $z_i$ is very important because it allows us to formulate the theory directly at $\lambda_M=0$ in terms of the fields $h_i$ and $z_i$.
In the following we set $H_{ext}=0$ for simplicity. The recursive equations for the fields are:
\beqa
h_0&=&\sum_{i=1}^{k+1}u(J_i,h_i)
\label{h0fin}
\\
z_0&=&\sum_{i=1}^{k+1}{du(J_i,h_i)\over dh_i}z_i
\label{z0fin}
\eeqa
The anomalous shift is given by the $\lambda_M\rightarrow 0$ limit of eq. (\ref{lambdaDM}): 
\beq
\Delta X^{(1)}=\sum_{i=1}^{k+1}\left(  \tanh \left(\sum_{j=1}^{k+1}\beta u (J_j,h_j)\right) -\tanh \beta u(J_i,h_i)       \right){d u \over d h_i} z_i
\label{dx1}
\eeq
The recursive equation for the distribution of the fields $h_0$ and $z_0$ is derived from eq. (\ref{Ph0h0p}):
\beqa
P_0(h_0,z_0)& =& 
\frac{1}{\int \prod_{i=1}^{k}P_i(h_i,z_i)dh_idz_i e^{u \beta \Delta F^{(iter)}+\Delta X^{(iter)}}} \int \prod_{i=1}^{k}P_i(h_i,z_i)dh_idz_i \times
\non
\\
& \times &
e^{u \beta \Delta F^{(iter)}+ \Delta X^{(iter)}} 
\delta \left(h_0-\sum_{i=1}^{k}u(J_i,h_i)\right)\delta\left(z_0-\sum_{i=1}^{k}{du(J_i,h_i)\over dh_i}z_i\right)
\label{Ph0z0}
\eeqa
The distribution of distributions ${\mathcal P}[P(h,z)]$ satisfies the following equation:
\beq
{\mathcal P}[P(h,z)]=\int \prod_{i=1}^k{\mathcal P}[P(h_i,z_i)]dP_i(h_i,z_i) \langle \delta( (P(h,z)-P(h_0,z_0))\rangle_J
\label{rec}
\eeq
where $P(h_0,z_0)$ depends on $P_i(h_i,z_i)$ through eq. (\ref{Ph0z0}) and the square brackets mean average over the $k$ bonds $J_i$.
The Complexity shift is
\beqa
\Delta \tilde{\Sigma}(u)^{(1)}& =&\ln \int \prod_{i=1}^{k+1}P_i(h_i,z_i)dh_idz_i\ \exp[u \beta \Delta F^{(1)}+\Delta X^{(1)}]
\label{dS1b}
\eeqa
According to equation (\ref{exte2}) the Legendre transform with respect to $u$ of the complexity is given by
\beq
\tilde{\Sigma}(u)={k+1\over 2}\int \prod_{i=1}^{k}{\mathcal P}(P_i(h_i,z_i))\langle \Delta \tilde{\Sigma}(u)^{(iter)}\rangle_J-{k-1\over 2}\int \prod_{i=1}^{k+1}{\mathcal P}(P_i(h_i,z_i))\langle \Delta \tilde{\Sigma}(u)^{(1)}\rangle_J
\label{comp1}
\eeq
where the square brackets mean average over the bonds $J_i$, and $\Delta \tilde{\Sigma}^{(iter)}$ has the same expression  $\Delta \tilde{\Sigma}^{(1)}$
with $k$ indexes instead of $k+1$.
At $u=0$ eq. (\ref{comp1}) gives the total complexity.

\section{Variational Expression of the Complexity}
\label{sec:var1}
In this section we derive the expression for the complexity that one can obtain considering the process of merging of $2k$ branches to a couples of spins according to eq. (\ref{exte1}). We will check that if the distribution of distributions ${\mathcal P}(P(h,z))$ satisfies equation (\ref{rec}) then the two expressions are equivalent. Furthermore it can be checked that the expression we recover is variational \cite{MP1,WS} in the sense that equation (\ref{rec}) 
 can be obtained extremizing it with respect to ${\mathcal P}(P(h,z))$. 

The free energy shift in the process of merging $2k$ branches with fields $\{h_1,\ldots,h_k\}$ and $\{g_1,\ldots,g_k\}$ to a couple of spin $\sigma_0$ and $\tau_0$ is \cite{MP1}:
\beqa
-\beta \D F^{(2)}&=&\sum_{i=1}^k
\ln \left[ {\cosh \beta J_i \over  \cosh \beta u(J_i,h_i)}{\cosh \beta K_i \over  \cosh \beta u(K_i,g_i)}\right]+
\non
\\
&+&\ln \left[ \sum_{\sigma_0 , \tau_0} \exp\left(    \beta J_0 \sigma_0 \tau_0 +\beta \sigma_0 \sum_{i=1}^k u(J_i,h_i)+\beta \tau_0 \sum_{i=1}^k u(K_i,g_i)  \right)   \right]
\label{df2}
\eeqa
Introducing the notation
\beq
y_i=\lambda_M {dg_i\over d H_{ext}} \ ,
\eeq
the corresponding anomalous shift can be obtained much as in the previous section:
\beqa
\D X^{(2)} & =& \sum_{i=1}^{k}\left[  \tanh \left(\sum_{j=1}^{k}\beta u (J_j,h_j)+\beta u\left(J_0, \sum_{j=1}^{k}\beta u (K_j,g_j)\right) \right) -\tanh \beta u(J_i,h_i)       \right]{d u \over d h_i} z_i+
\non
\\
& + & \sum_{i=1}^{k}\left[  \tanh \left(\sum_{j=1}^{k}\beta u (K_j,g_j) +\beta u\left(J_0, \sum_{j=1}^{k}\beta u (J_j,h_j)\right) \right) -\tanh \beta u(K_i,g_i)       \right]{d u \over d g_i} y_i
\label{dx2}
\eeqa
We call $Q_{i}(g_i,y_i)$  the distribution of the fields $g_i$ and $z_i$ to distinguish it from $P_i(h_i,z_i)$; then 
the Complexity shift reads
\beqa
\Delta \tilde{\Sigma}(u)^{(2)}& =&\ln \int \prod_{i=1}^{k}P_i(h_i,z_i)dh_idz_i \ Q_{i}(g_i,y_i)dg_idy_i \ \exp[u \beta \Delta F^{(2)}+\Delta X^{(2)}]
\eeqa
According to equation (\ref{exte1}) the Legendre transform with respect to $u$ of the complexity is given by
\beq
\tilde{\Sigma}(u)={k+1\over 2}\int \prod_{i=1}^{k}{\mathcal P}(P_i(h_i,z_i)){\mathcal P}(Q_{i}(g_i,y_i))  \langle \Delta \tilde{\Sigma}(u)^{(2)}\rangle_{J,K}-k\int \prod_{i=1}^{k+1}{\mathcal P}(P_i(h_i,z_i))\langle \Delta \tilde{\Sigma}(u)^{(1)}\rangle_J
\label{comp2}
\eeq
where the square brackets mean average over the bonds $J_i$ and $K_i$. Notice that at $u=0$ this expression gives the total complexity.
In order to prove the equivalence of eq. (\ref{comp1}) and eq. (\ref{comp2}) we need the following relationship that follows from the expressions of $\Delta F^{(2)}$, eq. (\ref{df2}) and of $\Delta F^{(1)}$, eq. (\ref{df1}) 
\beqa
&& \Delta F^{(2)}(h_1,\dots,h_k,g_1,\dots,g_k,J_1,\dots,J_k,K_1,\dots,K_k,J_0)=
\non
\\
&=& \Delta F^{(1)}(h_1,\dots,h_k,h_0,J_1,\dots,J_k,J_0)+F^{(iter)}(g_1,\dots,g_k,K_1,\dots,K_k)
\label{df2eq}
\eeqa
Where $h_0$ is the field generated by the $k$ fields $\{g_i\}$.
By derivation we obtain an analogous relationship between 
$\Delta X^{(2)}$, eq. (\ref{dx2}), and  $\Delta X^{(1)}$, eq. (\ref{dx1}):
\beqa
&& \Delta X^{(2)}(h_1,z_1,\dots,h_k,z_k,g_1,y_1\dots,g_k,y_k,J_1,\dots,J_k,K_1,\dots,K_k,J_0)=
\non
\\
&=& \Delta X^{(1)}(h_1,z_1,\dots,h_k,z_k,h_0,z_0,J_1,\dots,J_k,J_0)+\Delta X^{(iter)}(g_1,y_1,\dots,g_k,y_k,K_1,\dots,K_k)
\label{dx2eq}
\eeqa
Where $z_0$ is the field associated to $h_0$ and generated by the $k$ fields $\{g_i,y_i\}$.
The distribution $P_0(h_0,z_0)$ can be obtained from the distributions $Q_{i}(g_i,y_i)$ through equation (\ref{Ph0z0}), using this equation and eqs. (\ref{df2eq}) and (\ref{dx2eq}) we obtain:
\beqa
&& \Delta \tilde{\Sigma}(u)^{(2)}(P_1(h_1,z_1),\dots,P_k(h_k,z_k),Q_1(g_1,y_1)\dots,Q_k(g_k,y_k),J_1,\dots,J_k,K_1,\dots,K_k,J_0)=
\non
\\
&=& \Delta \tilde{\Sigma}(u)^{(1)}(P_1(h_1,z_1),\dots,P_k(h_k,z_k),P_0(h_0,z_0),J_1,\dots,J_k,J_0)+
\non
\\
&+&\Delta\tilde{\Sigma}(u)^{(iter)}(Q_1(g_1,y_1),
\dots,Q_k(g_k,y_k),K_1,\dots,K_k)
\label{dS2eq}
\eeqa
Replacing this equation in eq. (\ref{comp2}), introducing a $\delta$-function on $P(h,z)$ like the one in eq. (\ref{rec}) and integrating over $P(h,z)$, we readily obtain the equivalence between expression (\ref{comp2}) and (\ref{comp1}).
In the following we prove that (\ref{comp2}) is a variational expression. We extremize it with respect to a generic distribution of distributions ${\mathcal P}[P(h,z)]$ with the condition that it is normalized. This can be done introducing a Lagrange multiplier that multiplies the integral of ${\mathcal P}[P(h,z)]$ over the space of the distributions $P(h,z)$; as a consequences the variational equations are:
\beq
{\delta \tilde{\Sigma} ({\mathcal P}[P(h,z)])  \over \delta {\mathcal P}[P(h,z)]}={\rm const.}
\eeq 
Deriving (\ref{comp2}) with respect to ${\mathcal P}[P(h,z)]$  we obtain:
\beqa
{\delta \tilde{\Sigma} ({\mathcal P}[P(h,z)])  \over \delta {\mathcal P}[P(h,z)]}&=&
\int \prod_{i=2}^{k}{\mathcal P}[P_i(h_i,z_i)]\times
\non
\\
&\times & k (k+1) \left( \int\prod_{i=1}^{k} {\mathcal P}[Q_{i}(g_i,y_i)]  \langle \Delta \tilde{\Sigma}(u)^{(2)}\rangle_{J,K}-\int  {\mathcal P}[P_0(h_0,z_0)] \langle \Delta \tilde{\Sigma}(u)^{(1)}\rangle_J
\right)
\label{devS}
\eeqa
Replacing eq. (\ref{dS2eq}) in the integrand and performing the same manipulations described above we can show that the integrand in the second line of (\ref{devS}) does not depend on ${\mathcal P}[P_i(h_i,z_i)]$, therefore it is a constant:
\beqa
{\delta \tilde{\Sigma} ({\mathcal P}[P(h,z)])  \over \delta {\mathcal P}[P(h,z)]}&={\rm const.}=&
 k (k+1) \int\prod_{i=1}^{k} {\mathcal P}[Q_{i}(g_i,y_i)]  \langle \Delta \tilde{\Sigma}(u)^{(iter)}\rangle_{K}
\eeqa

\section{The SK Limit}

In this section we will study the high connectivity limit of the theory recovering the BM calculation of the TAP complexity of the SK model \cite{BMan}.
The critical temperature satisfies the relationship:
\beq
\langle \tanh^2 \beta_c J\rangle_J={1\over k} 
\eeq
 in order to have $\beta_c=1$ at any $k$, and in particular in the high connectivity limit,
 the strength of the bonds must depend on $k$:
\beq
J=\pm \arctanh {1\over \sqrt{k}}=\pm {1\over \sqrt{k}}+O\left({1\over k}\right)\ \ \ , \ \ \beta_c=1
\eeq 
In this limit the functions $u$ and $du/dh$ entering the recursion equations eq. (\ref{h0fin}) and (\ref{z0fin}) become:
\beqa
u(J,h)&=&J \tanh \beta h 
\\
{du \over dh}&=&J \beta (1- \tanh^2 \beta h)
\eeqa
As a consequence the recursion equations  (\ref{h0fin}) and (\ref{z0fin}) become:
 \beqa
h_0&=&\sum_{i=1}^{k}J_i \tanh \beta h_i 
\label{h0finSK}
\\
z_0&=&\sum_{i=1}^{k}J_i \beta (1- \tanh^2 \beta h_i)
z_i
\label{z0finSK}
\eeqa
The anomalous shift of the iterating process is the analogue of  eq. (\ref{dx1}):
\beq
\Delta X^{(iter)}=\tanh (\beta h_0)z_0- \sum_{i=1}^{k}\tanh \beta u(J_i,h_i)       {d u \over d h_i} z_i
\eeq
where we have replaced the recursive equations for $h_0$ and $z_0$ in the first term. In the high connectivity limit the second term becomes a constant and we have:
\beq
\Delta X^{(iter)}=\tanh (\beta h_0)z_0+{\rm const.}
\label{dxSK}
\eeq
The distribution function of the variables $h_0$ and $z_0$ before the reweighing is a Gaussian, see \cite{MPV}, Pg. 71. In principle the means of this Gaussian distribution depend on the disorder; for simplicity however we assume that this is not the case. This hypothesis yields the annealed complexity, (see discussion in \cite{R}) and is correct for the total complexity according to the BM theory \cite{BMan}.
According to eqs. (\ref{h0finSK}) and (\ref{z0finSK}) the covariances of the Gaussian distribution are
\beqa
\langle h_0^2\rangle_{Gauss} & = & {1\over k}\sum_{i=1}^{k} \tanh^2 \beta h_i= \langle m^2\rangle=q 
\label{g1}
\\
\langle h_0 \, z_0\rangle_{Gauss} & = & {1\over k}\sum_{i=1}^{k}\beta  \tanh \beta h_i (1- \tanh^2 \beta h_i)z_i=\beta \langle m (1-m^2)z\rangle={\Delta\over \beta}
\label{g2}
\\
\langle z_0^2 \rangle_{Gauss} & = & {1\over k}\sum_{i=1}^{k}\beta^2  (1- \tanh^2 \beta h_i)^2z_i^2=\beta^2 \langle (1-m^2)^2z^2\rangle=2 \lambda
\label{g3}
\eeqa
The averages labelled by the suffix {\it Gauss.} refer to the distribution before the reweighing. The averages without suffix refer to the reweighed distribution we will derive below. The last relation on each line is respectively the definition of the parameter $q$, $\lambda$ and $\Delta$.
According to eq. (\ref{dxSK}) the reweighted distribution is:
\beq
P(h,z)=K \exp[-{1\over 2} \left(\begin{array}{cc} h & z \end{array}\right)C^{-1}\left(\begin{array}{c} h \\ z \end{array}\right)+z \tanh\beta h]
\label{rewhz}
\eeq
where
\beq
C=\left(\begin{array}{cc} q & {\Delta \over \beta}\\ {\Delta \over \beta}& 2 \lambda \end{array}\right)
\eeq
and $K$ is the normalization constant.
Performing the change of variable 
\beq
m=\tanh \beta h\ ,
\label{m}
\eeq
 and integrating over $z$, we obtain the BM form \cite{BMan}:
\beq
P(m)=K' \left({1\over 1-m^2}\right)\exp\left[ -{ (\tanh^{-1}m- \Delta m)^2 \over 2 q \beta^2}+\lambda m^2\right]dm
\label{rewm}
\eeq
Where $K'$ is a normalization constant.
The expression of the reweighed distribution can be used to compute self-consistently the parameters $q$, $\lambda$ and $\Delta$. According to eqs. (\ref{g1},\ref{g2},\ref{g3}), we have:
\beqa
q&=&\langle m^2\rangle 
\label{g1f}
\\
{\Delta\over \beta}&
=&\beta \langle m (1-m^2)z\rangle
\label{g2f}
\\
2 \lambda
&=&\beta^2 \langle (1-m^2)^2z^2\rangle\label{g3f}
\eeqa
Where the square brackets mean average with respect to the reweighed distribution eq. (\ref{rewhz}).
Integrating over $z$ we immediately recover from eq. (\ref{g1f}) the analogous equation of the BM theory \cite{BMan}.
In order to simplify the other equations we notice that:
\beq
\beta (1-m^2)z={d \over dh}z \tanh \beta h
\label{dtan}
\eeq
Introducing the shorthand notation $G(h,z)$ for the Gaussian part of $P(h,z)$:
\beq
G(h,z)=\exp[-{1\over 2} \left(\begin{array}{cc} h & z \end{array}\right)C^{-1}\left(\begin{array}{c} h \\ z \end{array}\right)
\eeq
we have:
\beqa
 {\Delta \over \beta}&
 = & K \int m G(h,z){d \over d h}\exp[z \tanh \beta h]
]dhdz=
\non
\\
&=& K \int -{dm\over dh}G(h,z)\exp[z \tanh \beta h]dh dz-K\int m \left( {d\over dh}G(h,z) \right)\exp[z \tanh \beta h]dh dz=
\non
\\
&=& \beta (1-q)-K\int m \left( {d\over dh}G(h,z) \right)\exp[z \tanh \beta h]dh dz
\eeqa
The integration over $z$ in the second term can be performed explicitly, see \cite{R}, the final result is equal to the BM equation for $\Delta$:
\beq
\Delta=-{\beta^2 \over 2}(1-q)+{1\over 2 q}\langle m \tanh^{-1}m \rangle
\eeq
The equation for $\lambda$ can be derived analogously; from eq. (\ref{g3f}) and eq. (\ref{dtan}) we have
\beqa	
2 \lambda &=& K \int G(h,z) \left({d\over dh}z \tanh\beta h   \right)^2\exp[z \tanh \beta h]dh dz
=
\non
\\
&=&
K\int  \left( {d^2\over dh^2}G(h,z) \right)\exp[z \tanh \beta h]dh dz+
K\int  G(h,z){d^2 \over d h^2}\exp[z \tanh \beta h]
]dhdz
\eeqa
The integration over $z$ in the first term can be performed explicitly, following \cite{R} we have: 
\beq
K\int  \left( {d^2\over dh^2}G(h,z) \right)\exp[z \tanh \beta h]dh dz=-{1\over q}\left( 1- \frac{\langle (\tanh^{-1}m-\Delta m)^2 \rangle}{q \beta^2}\right)      
\eeq
To compute the second term we note that:
\beq
{d^2 \over dh^2}z \tanh \beta h= 2 \beta^2 z m (1-m^2)
\label{d2tan}
\eeq
Therefore, according to the equation for $\Delta$, eq. (\ref{g2f}) we have:
\beq
K\int  G(h,z){d^2 \over d h^2}\exp[z \tanh \beta h]
=2 \Delta
\eeq
Putting all together we obtain 
\beq
\lambda= \Delta-{1\over 2 q}\left( 1- \frac{\langle (\tanh^{-1}m-\Delta m)^2 \rangle}{q \beta^2}\right)     
\eeq
This is precisely the equation corresponding to the extremization of the BM action  with respect to the parameter $q$.

\section{The Annealed Theory}
\label{sec:ann}

In this section we develop the annealed formulation of the theory. The advantage of this approach is that it gives non-trivial information on the complexity at the price of the replica-symmetric solution, {\it i.e.} considering a single distribution $P_{ann}(h,z)$.

We discuss the annealed formulation considering the standard cavity method with only the field $h$ and the parameter $u$. The generalization to the present theory is straightforward and we will simply report the full equations at the end of this section.
Instead of computing the Legendre transform of the physical complexity, {\it i.e.} 
\beq
\tilde{\Sigma}(u)=\overline{\ln \sum_{\alpha}e^{-u \beta f_{\alpha}}}\, ,
\eeq
where the bar means average over the disorder,
we compute 
\beq
\tilde{\Sigma}_{ann}(u)=\ln \overline{\sum_{\alpha}e^{-u \beta f_{\alpha}}}.
\eeq
Notice that although the correct physical object is the first one, the second one is a well-defined analytical object that gives an upper bound to the first because of the convexity of the logarithm.
It is easy to verify that this object is the Legendre transform of the annealed complexity defined as:
\beq
\Sigma (f)=\ln \overline{\sum_{\alpha}\delta (f-f_{\alpha})}
\eeq
The derivatives of the annealed complexity produce annealed averages defined as:
\beq
\langle O \rangle = \frac{\overline{\sum_{\alpha}O_{\alpha}e^{-u \beta f_{\alpha}}} } {\overline{\sum_{\alpha}e^{-u \beta f_{\alpha}}}}\,  .
\eeq
The quantity $\tilde{\Sigma}(u)$ is clearly an extensive quantity; therefore, according to section \ref{subsec:theext}, it can be expressed as the sum of proper shifts, see eqs. (\ref{exte1}) and (\ref{exte2}). Thus we are interested in the shift in the process of merging $k+1$ branches that reads:
\beq
\Delta \tilde{\Sigma}^{(1)}_{ann}(u)=\ln\overline{\sum_{\a}e^{u \beta F_{\a}+u \beta \Delta F_{\a}^{(1)}}}-\ln\overline{\sum_{\a}e^{u \beta F_{\a}}}=\ln\frac{\overline{\sum_{\a}e^{u \beta F_{\a}+u \beta \Delta F_{\a}^{(1)}}}}{\overline{\sum_{\a}e^{u \beta F_{\a}}}}
\eeq
According to the hypotheses of the cavity method, on each sample the free energies are not correlated with the fields and with the free energy shift, see the discussion of eq. (\ref{grancav}), therefore we may write:
\beq
\sum_{\a}e^{u \beta F_{\a}+u \beta \Delta F_{\a}^{(1)}}=\sum_{\a}e^{u \beta F_{\a}}\int \prod_{i=1}^{k+1}P_i{h_i}dh_i \exp[\beta u \Delta F^{(1)}(J_1,\dots,J_k,h_1,\dots,h_k)]
\label{expshift}
\eeq
In order to perform the disorder average of this quantity we note that before the merging process the branches are not correlated and the states of the system of $k+1$ branches are the product of the states of each branch with free energy equal to the sum of the free energy on each branch.
Thus we may write the following  relationships:
\beq
\sum_{\alpha}e^{\beta u F_{\alpha}}=\prod_{i=1}^{k+1}\sum_{\alpha_i}e^{\beta  u F_{\alpha_i} }
\eeq
\beq
\overline{\sum_{\alpha}e^{\beta u F_{\alpha}}}=\prod_{i=1}^{k+1}\overline{\sum_{\alpha_i}e^{\beta  u F_{\alpha_i} }}
\label{fat}
\eeq
\beq
\overline{\sum_{\alpha}e^{\beta u F_{\alpha}}\prod_{i=1}^{k+1}P_i(h_i)}=\prod_{i=1}^{k+1}\overline{\sum_{\alpha_i}e^{\beta  u F_{\alpha_i} }P_i(h_i)}
\eeq
as a consequence the disorder average of eq. (\ref{expshift}) reads
\beq
\overline{\sum_{\a}e^{u \beta F_{\a}+u \beta \Delta F_{\a}^{(1)}}}=\int \prod_{i=1}^{k+1}\overline{\sum_{\alpha_i}e^{\beta  u F_{\alpha_i} }P_i(h_i)}
dh_i \langle \exp[\beta u \Delta F^{(1)}(J_1,\dots,J_k,h_1,\dots,h_k)] \rangle_{J}
\label{dsnum}
\eeq
We introduce the function $P_{ann}(h)$ defined as the annealed average of the field distribution $P_i(h_i)$ of the generic branch $i$
\beq
P_{ann}(h)=\frac{\overline{\sum_{\alpha_i}e^{\beta  u F_{\alpha_i} }P_i(h_i)}}{\overline{\sum_{\alpha_i}e^{\beta  u F_{\alpha_i} }}}\, ,
\eeq
then replacing in  eq. (\ref{dsnum}) and using eq. (\ref{fat}) we obtain:
\beq
\Delta \tilde{\Sigma}_{ann}^{(1)}(u)=\ln\int \prod_{i=1}^{k+1}P_{ann}(h_i)dh_i \langle \exp[\beta u \Delta F^{(1)}(J_1,\dots,J_k,h_1,\dots,h_k)] \rangle_{J}
\label{DS1annb}
\eeq
The recursive equation for the annealed distribution $P_{ann}(h)$ can be obtained in a similar way starting from the iterative equation on a give sample, eq. (\ref{Ph0z0}).
The annealed theory can be easily generalized to the case with non-zero fields $z$ and non-zero anomalous shifts $\Delta X$. The complexity shifts are:
\beq
\Delta \tilde{\Sigma}_{ann}^{(1)}(u)=\ln\int \prod_{i=1}^{k+1}P_{ann}(h_i,z_i)dh_i\, dz_i \langle \exp[\beta u \Delta F^{(1)}+\Delta X^{(1)}] \rangle_{J}
\eeq
\beq
\Delta \tilde{\Sigma}_{ann}^{(2)}(u)=\ln\int \prod_{i=1}^{k}P_{ann}(h_i,z_i)dh_i\,dz_i P_{ann}(g_i,y_i)dg_i\,dy_i\langle \exp[\beta u \Delta F^{(2)}+\Delta X^{(2)}] \rangle_{J,K}
\label{DS2ann}
\eeq
where the free energies and anomalous shifts where defined in eqs. (\ref{df1},\ref{dx1},\ref{df2},\ref{dx2}).
The Legendre transform of the complexity can be obtained from eqs. (\ref{exte1}) or (\ref{exte2}) as:
\beqa
{\tilde{\Sigma}_{ann}\over N}&=&{k+1\over 2}\Delta \tilde{\Sigma}_{ann}^{(2)}(u)-k\Delta \tilde{\Sigma}_{ann}^{(1)}(u)
\label{S2var}
\\
{\tilde{\Sigma}_{ann}\over N}&=&{k+1\over 2}\Delta \tilde{\Sigma}_{ann}^{(iter)}(u)-{k-1\over 2}\Delta \tilde{\Sigma}_{ann}^{(1)}(u)
\label{S1var}
\eeqa
The equation for $P_{ann}(h,z)$ is:
\beqa
P_{ann}(h,z)& =& 
\frac{1}{\int \prod_{i=1}^{k}P_{ann}(h_i,z_i)dh_idz_i \langle e^{u \beta \Delta F^{(iter)}+\Delta X^{(iter)}}\rangle_J} \int \prod_{i=1}^{k}P_{ann}(h_i,z_i)dh_idz_i \times
\non
\\
& \times &
\langle e^{u \beta \Delta F^{(iter)}+ \Delta X^{(iter)}} 
\delta \left(h-\sum_{i=1}^{k}u(J_i,h_i)\right)\delta\left(z-\sum_{i=1}^{k}{du(J_i,h_i)\over dh_i}z_i\right) \rangle_J 
\label{Phzann}
\eeqa

\subsection{Variational Formulation of the Annealed Theory}
The annealed theory admits a variational formulation much as the general theory. Indeed in  this section we will show that the equation (\ref{Phzann}) is identical with the equation one obtains  extremizing expression (\ref{S2var}) with respect to the function $P_{ann}(h,z)$.  Notice that while in the general theory the object over which we extremize is the distribution of distributions ${\mathcal P}[P(h,z)]$, in the annealed theory the object over which we extremize is simply $P_{ann}(h,z)$. 
In order to prove this property we start by proving the equivalence of the expressions (\ref{S2var}) and (\ref{S1var}). To do this we recall the relationships that connect the free energy and anomalous shift in various systems, eq. (\ref{df2eq}) and eq. (\ref{dx2eq}). Putting these equations in eq. (\ref{DS2ann}) and introducing two $\delta$-function over $h_0$ and $z_0$ like those of eq. (\ref{Phzann}) as in section \ref{sec:var1} we obtain:
\beq 
\Delta \tilde{\Sigma}_{ann}^{(2)}=\Delta \tilde{\Sigma}_{ann}^{(1)}+\Delta \tilde{\Sigma}_{ann}^{(iter)}
\label{equiv}
\eeq
This equation yields immediately the equivalence between the expressions (\ref{S2var})
and (\ref{S1var}).
The functional derivative of expression (\ref{S2var}) with respect to $P_{ann}(h,z)$ is
\beq
{\delta \tilde{\Sigma}\over \delta P_{ann}(h,z)}=k (k+1)\int\prod_{i=2}^k P_{ann}(h_i,z_i)dh_i\,dz_i\,I
\eeq
Where the integrand $I$ is given by:
\beqa
I&=& e^{-\Delta \tilde{\Sigma}_{ann}^{(2)}}\int \prod_{i=1}^{k}P_{ann}(g_i,y_i)dg_i\,dy_i\langle \exp[\beta u \Delta F^{(2)}+\Delta X^{(2)}] \rangle_{J,K}+
\non
\\
&-&e^{-\Delta \tilde{\Sigma}_{ann}^{(1)}} \int P_{ann}(h_0,z_0)dh_0 dz_0 \langle \exp[\beta u \Delta F^{(1)}+\Delta X^{(1)}] \rangle_{J}
\eeqa
Using once again eq. (\ref{df2eq}) and eq.(\ref{dx2eq}) and introducing two $\delta$ functions on $h_0$ and $z_0$ like those appearing in eq. (\ref{Phzann}) we obtain:
\beq
I=(e^{-\Delta \tilde{\Sigma}_{ann}^{(2)}+\Delta \tilde{\Sigma}_{ann}^{(iter)}}-e^{-\Delta \tilde{\Sigma}_{ann}^{(1)}})\int P_{ann}(h_0,z_0)dh_0 dz_0 \langle \exp[\beta u \Delta F^{(1)}+\Delta X^{(1)}] \rangle_{J}=0
\eeq 
the latter equality follows from eq. (\ref{equiv}) and proves that expression (\ref{S2var}) is extremized by $P_{ann}(h,z)$.

\section{Solving the Equations}
\label{sec:num}

We have studied the annealed and the general theory at $u=0$ by means of population dynamics algorithms. 
The most important result of this analysis is that in both cases the algorithms converge to  populations with  non-zero values of the fields $z$ meaning that the whole theory is non trivial.

The procedure is the same of M\'ezard and Parisi \cite{MP1}: thereby we consider a population of ${\mathcal N}$ populations of ${\mathcal M}$ couples of fields $\{h,z\}$ and we replace one of  its elements generating a new population. The replacement is done sequentially over the  $\mathcal N$ populations.
To generate the populations
we used a simplified version of algorithm B of \cite{MP1}, that is at each iteration we generate through a merging procedure a population of ${\mathcal M}$ couples of fields $ \{h,z\}$ with the corresponding $\Delta X^{(iter)}$ 
shifts, then we generate a new population picking up ${\mathcal M}$ couples of fields from the original population with weights proportional to $\exp[\Delta X^{(iter)}]$.
The main difference with the standard case is that we have to consider couples of fields instead of the single cavity field, therefore in order to obtain the same precision on the function $P(h,z)$ we need to consider larger values of ${\mathcal M}$ with larger computation time.

\begin{figure}[htb]
\begin{center}
\epsfig{file=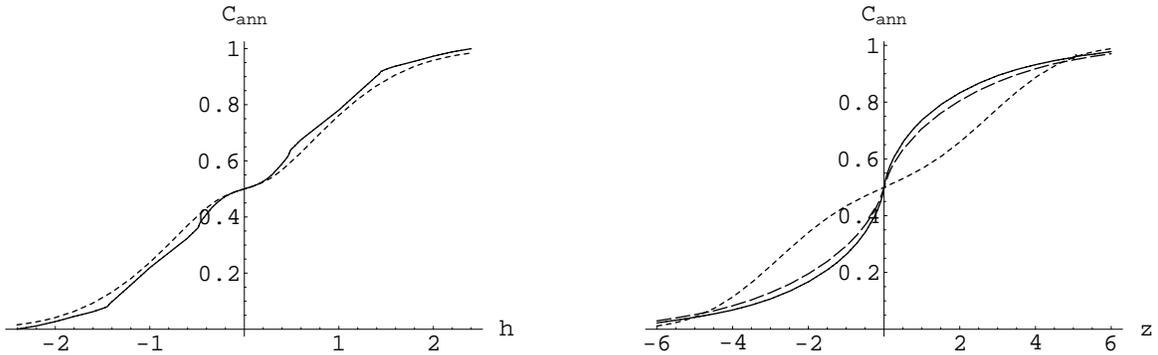,width=17cm} 
\caption{Left plot, continuous line: the function $C_{ann}(h)$ defined in the text, at $k=5$, $T=.3$; dashed line: the function $C_{ann}(h)$ in the SK model at the same temperature.
Right plot, continuous line: the function $C_{ann}(z)$ defined in the text, at $k=4$, $T=.3$; long-dashed line: the function $C_{ann}(z)$ at $k=5$, $T=.3$;  dashed line: the function $C_{ann}(z)$ in the SK model at the same temperature}
\label{figure1}
\end{center}\end{figure}

The annealed theory requires only one population and thus more precise results can be obtained.
In figure \ref{figure1} we plot the cumulative functions with respect to $h$ and $z$ of the function $P_{ann}(h,z)$ at temperature $T=.3$ and connectivity $k=4,5$:
\beq
C_{ann}(h)=\int_{-\infty}^{\infty}dz\int_{h_{min}}^h\, dh \,P_{ann}(h,z)
\eeq 
\beq
C_{ann}(z)=\int_{-\infty}^{z}dz\int_{h_{min}}^{h_{max}}\, dh \,P_{ann}(h,z)
\eeq 
where according to eq. (\ref{h0fin}) we have
$
{h_{max}}=-{h_{min}}=k \,\arctanh (k^{-1/2}). 
$ 
 The dashed lines represent the corresponding SK result given by eq. (\ref{rewhz}). 

\begin{figure}[htb]
\begin{center}
\epsfig{file=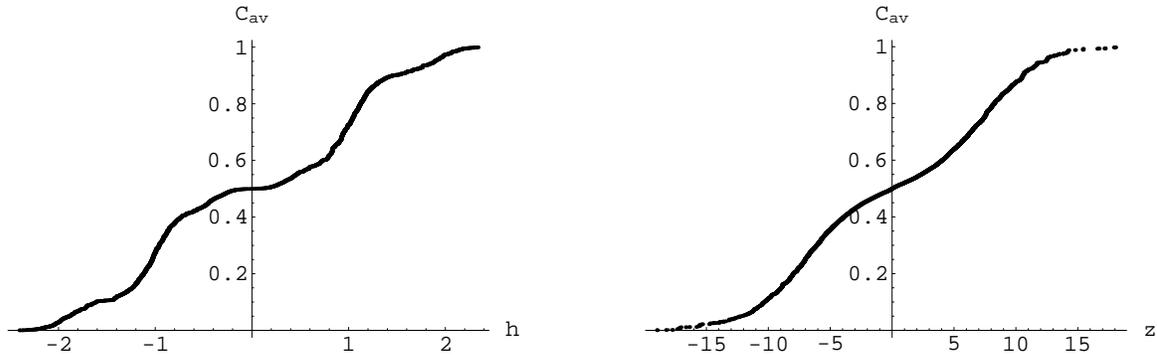,width=17cm} 
\caption{Left plot: the cumulative function of an average population representing the function $C_{av}(h)$ defined in the text, at $k=5$, $T=.3$ obtained from a population of ${\mathcal N}=300$ populations of ${\mathcal M}=1024$ couples of fields $\{h,z\}$. Right plot; the cumulative function of an average population representing the function $C_{av}(z)$ defined in the text, at $k=5$, $T=.3$ obtained from the same  population of ${\mathcal N}=300$ populations.}
\label{figure2}
\end{center}\end{figure}

In figures (\ref{figure2})
we plot the cumulative functions $C_{av}(h)$ and $C_{av}(z)$ of the quenched average $P_{av}(h,z)$ of the distributions $P_i(h_i,z_i)$:
\beq
P_{av}(h,z)=\int P(h,z)
{\mathcal P}[P(h,z)]dP(h,z)
\eeq
The function $P_{av}(h,z)$ in the plot is represented by an average population of ${\mathcal M N}$ couples  obtained merging in a single population all the ${\mathcal N}=300$ populations of ${\mathcal M}=1024$ couples of fields at temperature $T=.3$ and connectivity $k=5$.
The function $P_{av}(h,z)$ must be a smooth function invariant under the exchange  of $\{h,z\}$ with $\{-h,-z\}$, but this is true only approximately for the cumulative distributions of the average population plotted in figure (\ref{figure2}) 
 due to the finite dimension ${\mathcal N}$ of the population of populations; indeed the single populations are not invariant  with respect to the exchange of $\{h,z\}$ with $\{-h,-z\}$ and  have quite large deviations  from $P_{av}(h,z)$. 

The complexity is extremely small at finite temperature, therefore its value and in general the shape of the function $\Sigma(f)$ require a more refined treatment that goes beyond the scope of this work.
To confirm the smallness of the complexity we have studied the annealed theory by means of computer assisted series expansion in power of the reduced temperature $\tau=T_c-T$. 
These expansions yield good results near $T_c=1$ and can be used in the whole low temperature phase provided some resummation scheme is applied \cite{CR,noiquen}. At $k=2$ we found 
\beq
\Sigma_{ann}(k=2)=2 \left(\arctanh {1 \over \sqrt{2}}\right)^6 \tau^6+O(\tau^7)
\eeq
we note that in the  SK model one finds the same $\tau^6$ behaviour \cite{BMan,CGPM,noiquen}.

\section{Discussion}

We have obtained a theory for the Complexity of the Bethe lattice spin-glass which is the analogue of the non-SUSY solution for the TAP complexity of the SK model.
As explained in section \ref{sec:bas}, within this theory the relevant states are marginal, a feature that is encoded in the finiteness of the rescaled fields $z$ or equivalently in the divergence of the fields $h'$. 
Fixing a finite $\lambda_M$ we find a  set of states with a finite complexity lower than the total one, the crucial point is that these states are no more marginal and the cavity method can be applied to them. The total complexity is obtained as the $\lambda_M\rightarrow 0$ limit of the complexity of these states. 
To select a set of states different from the relevant ones we have considered states with a magnetization different from that of the relevant states, while in \cite{R} the selfoverlap was used.
In the present context we could have considered any parameter $A$, different from the free energy, that can be written as
\beq
A=-{dF\over d\tilde{A}} \, ,
\eeq  
where $\tilde{A}$ is the conjugated field of the observable $A$, not to be confused with the parameter $\lambda_A$ that weighs the states. For instance we could have considered the entropy instead of the magntetization with the temperature as conjugated field. Here we want to show that in the limit $\lambda_A\rightarrow 0$ we would have obtained the same result for the total complexity and in general for the curve $\tilde{\Sigma}(u)$. The proof is simple, indeed the shift of the parameter $A$ can be obtained deriving the free energy shift and has two terms much as eq. (\ref{dM1}):
\beq
\Delta A=-{d \Delta F\over d\tilde{A}}=-{\partial \Delta F\over \partial \tilde{A}}-\sum_i{\partial \Delta F\over \partial h_i}{d h_i\over d \tilde{A}}
\label{dA}
\eeq
Now if we define the fields $z$ as
\beq
z_i=\lambda_A{d h_i\over d \tilde{A}}
\eeq
It is easily seen that 
\beq 
\lim_{\lambda_A\rightarrow 0}\lambda_A \Delta A=\Delta X
\eeq
Indeed the first term in (\ref{dA}) does not depend on the $z$ and vanishes anyway. This is the only term whose structure depends explicitly on the parameter $A$, instead the second term is simply proportional to the derivatives with respect to  the fields $h_i$ and the dependence on $A$ is hidden in the definition of the fields $z_i$.

In section \ref{sec:ann} we have presented the annealed version of the theory. The annealed formulation has the advantage of yielding non-trivial results requiring only one distribution, {\it i.e.} we can study the dynamics of a single population. At variance with the RS result which is wrong physically and analitically, the annealed complexity is a well-defined object although the typical complexity is the quenched one yielded by the general theory. In general it gives an upper bound to the real complexity. 

We notice that the annealed theory coincides with the so-called factorized approximation \cite{WS,goldlai,MP2}. Thus the factorized approximation has always a well-defined analytical meaning. 
In \cite{MP2,MPR} it has been pointed out that in the $T\rightarrow 0$ limit the field distributions should become concentrated over the integers and it has been argued that this property can be used to test wether a 1RSB solution is correct or is an approximation to a solution with more steps of replica-symmetry breaking.
Thus an interesting question is wether the annealed solution satisfies this property. We note {\it en passant} that in the SK model the annealed solution is alway described by the BM solution {\it i.e.} the $z$ fields are always non-zero, even at the lower band-edge which does not coincide with the equilibrium solution.

In the SK model the quenched and the annealed total complexities coincide, and this remains true for the states of a given free energy $f$, provided $f$ is greater than some $f_c$ \cite{BMan}. Instead here we found that the annealed solution for the total complexity is unstable with respect to 1RSB and one may ask if the 1RSB solution is stable towards 2RSB and wether there is a given value of the free energy such that the annealed solution is stable towards 1RSB.

The Ising $p$-spin model exhibits a phase transition from a non-SUSY complexity curve at high free energies similar to the one of the SK model \cite{rieger} to a SUSY solution at low free energies that ends at the 1RSB equilibrium free energy \cite{MR,isp,R}. The theory presented here can be extended to 
 other models with finite connectivity such as XORSAT \cite{FLRZ} and $k$-SAT \cite{MZ,MZP}and it would be interesting to check for the presence of such a transition which would be marked by the presence of non-zero $z$ fields possibly leading to different thresholds.
We also mention that a stability condition for the solution with non-zero $z$ analogous to the condition of positivity of the replicon \cite{MPR} is currently under investigation. Indeed in the Ising $p$-spin model the transition from a SUSY solution to a non-SUSY solution occurs precisely where the replicon eigenvalue becomes negative \cite{isp}.  
Thus to cure this instability it is sufficient to consider non-zero $z$-field without considering higher steps of RSB as proposed in \cite{MR}.

In section \ref{sec:num} we presented an investigation of the theory at finite temperature. 
We have checked that population dynamics algorithms converge to non-trivial solutions of the equations of the theory, in particular the fields $z$ are non-zero. A detailed analysis of the resulting complexity curves would require higher computational efforts because the complexity is extremely small at finite temperature. Indeed through series expansion in power of $\tau=T_c-T$ near $T_c=1$ we have checked that the behaviour of the total annealed complexity  is the same of the SK model, {\it i.e.} $\Sigma=O(\tau^6)$ \cite{BMan}.
It seems reasonable to expect that the behaviour of the complexity curve $\Sigma(f)$ be the same  as the BM solution of the SK model, {\it i.e.} a bell-shaped curve markedly different from the supersymmetryc (SUSY) solution \cite{PaSou1,PP,CGPM,noian} that instead corresponds to set $z=0$ at all values of $u$ \cite{MP2,Moncomp}.
However, as noted in \cite{noian}, the lower band edge of the curve $\Sigma(f)$ in SK is described  by the SUSY solution and therefore, as far as equilibrium properties are concerned, we do not need to consider the fields $z$ because they vanish. Thus at the lower band edge the full-RSB version of the theory becomes the same of the full-RSB version of the  M\'ezard and Parisi theory \cite{MP1}. 
Numerically we can study only  approximations of the real FRSB distribution, and
in principle the theory with non-zero $z$ may give a prediction for the equilibrium free energy $f_{eq}$ worse than the one of \cite{MP1} that is in extremely good agreement with the numerical results quoted in \cite{MP1}. 
The point deserves more investigation also in the SK model, indeed in SK the 1RSB SUSY solution yields a better result for $f_{eq}$ than the annealed solution but the properties of the quenched BM solution are not known at the present moment, neither at 1RSB nor at FRSB.

In view of the extension to other finite connectivity models and optimization problems,  it is very interesting to study the theory in the $T\rightarrow 0$ limit and exactly at $T=0$, work is in progress in these directions.
Here we note that at low temperature the fields $z$ diverge in a non-trivial way in the SK model. 
Indeed the parameters $\Delta$ and $\lambda$ governing the BM theory diverge linearly with $\beta$ with the same prefactor, as a consequence the Gaussian distribution of $z$ and $h$ before the reweighting eqs. (\ref{g1},\ref{g2},\ref{g3}) has the following behaviour at low $T$:
\beqa
\langle h_0^2\rangle_{Gauss} & = & q \simeq 1 
\label{g1T0}
\\
\langle h_0 \, z_0\rangle_{Gauss} & = & {\Delta\over \beta}=O(1)
\label{g2T0}
\\
\langle z_0^2 \rangle_{Gauss} & = & \lambda=O(\beta)
\label{g3T0}
\eeqa
thus the $z$ fields diverge in a way that cannot be accounted for through a simple rescaling while the $h$ fields remain finite. 

{\bf Acknowledgements:} It is a pleasure to thank G. Parisi for valuable advice and all the people at the SMC center who supported me during the preparation of this work.

\end{document}